\documentclass[11pt]{article}
\usepackage{moriond,epsfig}

\def\Journal#1#2#3#4{{#1} {\bf #2}, #3 (#4)}

\def\NPB{{\em Nucl. Phys.} B}

\def\PRD{{\em Phys. Rev.} D}

\def\be{\begin{equation}}
\def\ee{\end{equation}}
\def\bea{\begin{eqnarray}}
\def\eea{\end{eqnarray}}

\newcommand{\ri}{{\rm i}}
\newcommand{\e}{{\rm i}\epsilon}
\newcommand{\nn}{\nonumber}
\newcommand{\eir}{\varepsilon}

\begin{document}
\vspace*{4cm}
\title{NLO CALCULATIONS OF THE EXCLUSIVE PROCESSES IN PQCD}

\author{ \underline{G. DUPLAN\v CI\' C}, B. NI\v ZI\' C}

\address{Theoretical Physics Division, Rudjer Bo\v{s}kovi\'{c} Institute,\\ 
        P.O. Box 180, HR-10002 Zagreb, Croatia}

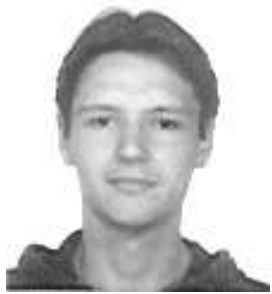
\maketitle\abstracts{We present a generally applicable reduction formalism 
which makes it possible to express an arbitrary tensor and scalar 
one-loop Feynman integral, with N external lines and massless propagators, 
in terms of a basic set of eight fundamental scalar Feynman integrals 
with 2, 3, and 4 
external lines, for arbitrary external kinematics.
The formalism is particularly suitable for the 
NLO calculations of
exclusive processes at large momentum transfer in pQCD, 
where all previously developed reduction 
methods fail due to the presence of collinear external
on-shell lines. 
}

\section{Motivation}

In testing various aspects of QCD the hadronic
exclusive processes (EP) at large momentum transfer 
in which the total number of particles (partons) in the initial and final
states is $N\ge 6$ are becoming increasingly important.

Leading order (LO) predictions have been obtained for many EP processes.
Owing to the 
fact that the LO predictions in
perturbative QCD (pQCD) do not have much predictive power, the
inclusion of higher-order corrections is essential because
they have a stabilizing 
effect,
reducing the dependence of the LO predictions on the
renormalization and factorization scales and the 
renormalization scheme. However, only a few EP have
been analyzed at the next-to-leading order (NLO). 

Obtaining radiative corrections in pQCD requires the evaluation
of one-loop Feynman integrals with massless propagators
(quarks masses can be neglected in high-energy processes).
These integrals contain 
IR divergences (both soft and collinear) and
need to be regularized. The most suitable regularization method for
pQCD calculations is dimensional regularization.

As it is well known, in calculating Feynman diagrams 
mainly three difficulties arise:
reduction of tensor integrals to scalar integrals, reduction of scalar
integrals to a set of basic scalar integrals and the evaluation
of the basic scalar integrals. 

Considerable progress has recently been made in developing
efficient approaches for calculating one$-$loop Feynman
integrals with a large number of external lines.
As far as the calculation of one-loop $N$-point
massless integrals is concerned,
the most complete and systematic method has been developed
by Binoth at al \cite{binoth}. It does not, however, apply to all 
cases of practical interest. Namely, being obtained for the
non-exceptional external momenta it cannot be applied
to the integrals in which the set of external momenta
contains subsets comprised of two or three collinear
on-shell momenta. 
Integrals of this type arise when
performing the leading-twist NLO analysis of hadronic
EP at large momentum transfer in pQCD. 

With no restrictions regarding the external kinematics,
in this paper we describe an efficient, systematic and
completely general method for reducing an arbitrary one$-$loop
$N-$point massless integral to a set of basic integrals.

\section{Reduction of tensor integrals}

In order to obtain one$-$loop radiative corrections
to physical processes in massless gauge theory,
the integrals of the following type are required:
\begin{equation}
I^N_{\mu _1\cdots \mu_P}(D;\{{\nu}_i\})=
(\mu ^2)^{2-D/2}\int
\frac{{\rm d}^D l}{(2\pi)^D}
\frac{l_{\mu_1}\cdots l_{\mu_P}}{A_1^{{\nu}_1}
A_2^{{\nu}_2} \cdots
A_N^{{\nu}_N}}, \qquad A_i = (l+r_i)^2+\e . \label{f4} 
\end{equation}
This is a rank $P$ tensor one-loop N-point Feynman 
integral with massless internal lines in $D$ dimensions,
where 
$l$ is the loop momentum, $\mu$ is the usual
dimensional regularization scale and ${\nu}_i\in \bf N$ are
arbitrary powers
of the propagators. 
The corresponding scalar integral, we denote by $I_0^N$.
These integrals represent generalizations of the 
usual integrals in practical calculations, where ${\nu}_i=1$.
However, the most natural presentation of the reduction method 
discussed here is in terms of these generalized integrals.

The corresponding expressions of the above integrals
in Feynman parameter space \cite{mi3} reads
\begin{eqnarray}
I^N_0(D;\{ \nu_i\})&=&\frac{
\ri }{(4\pi)^2}
(4\pi \mu ^2)^{2-D/2}\frac{\Gamma\left( 
\sum\nolimits_{i=1}^{N} \nu_i-D/2\right)}{
\prod\nolimits_{i=1}^{N} \Gamma
 (\nu_i)}(-1)^{\Sigma_{i=1}^N \nu_i}\int_0^1\! \left( \prod_{i=1}^{N}
 {\rm d}y_i 
y_i^{\nu_i-1}\right)\nn \\ 
& &\times\,
\delta\left(
\sum_{i=1}^{N}y_i-1\right)
\left[ \,-\sum_{i,j=1\atop i<j}^N y_i y_j 
\left( r_i-r_j\right)^2
-\e  \, \right]^{D/2-\Sigma_{i=1}^{N}
 \nu_i}\hspace{-1.5cm},\label{f13} \\
I^N_{\mu _1\cdots \mu_P}(D;\{ \nu_i\})&=& 
\sum_{k,
j_1,\cdots,j_{N}\geq 0\atop 2 k+{\scriptstyle
 \Sigma} j_i=P} \left \{ [g]^k [r_1]^{j_1}\cdots 
 [r_{N}]^{j_{N}}\right\}_{\mu _1\cdots \mu_P} 
 \frac{( 4 \pi \mu^2)^{P-k}}{(-2)^{k}} 
 \left[\prod_{i=1}^N \frac{\Gamma (\nu_i+j_i)}{
\Gamma(\nu_i)}\right]\nn \\
& &
{}\times I^N_0(D+2(P-k);\{ 
\nu_i+j_i\}), \label{f15} 
\end{eqnarray}
where $ \{ [g]^k [r_1]^{j_1}\cdots 
[r_{N}]^{j_{N}}\}_{\mu _1\cdots \mu_P}$
represents a symmetric (with respect to $\mu _1\cdots \mu_P$)
combination of tensors, each term of which is composed
of $k$ metric tensors and $j_i$ external momenta $r_i$
(for example,
$
\left \{ g r_1\right\}_{\mu _1 \mu_2 \mu_3}=
g_{\mu_1 \mu_2} r_{1\mu_3}+g_{\mu_1 \mu_3} r_{1\mu_2}+
g_{\mu_2 \mu_3} r_{1\mu_1}$).
The general results (\ref{f13}) and (\ref{f15})
represent massless versions of the results that have originally
been derived by Davydychev \cite{davy} for the case of massive 
Feynman integrals.
With the decomposition (\ref{f15}), the problem 
of calculating the tensor integrals has been reduced to the 
calculation of the general scalar integrals.

\section{Reduction of scalar integrals}
As is well known, the direct evaluation of the 
general scalar integral (\ref{f13}) 
represents a non-trivial problem.
However, with the help of recursion relations,
the problem can be significantly
simplified in the sense that the calculation of
the original scalar integral
can be reduced to the calculation of a certain number
of simpler fundamental (basic) integrals.

Owing to the translational invariance, 
the dimensionally
regulated integrals satisfy the following identity 
\cite{tarasov}:
\begin{equation}
0\equiv
\int
\frac{{\rm d}^D l}{(2\pi)^D}\frac{\partial}{\partial
l^\mu}\left(
\frac{z_0 l^{\mu}+\sum\nolimits_{i=1}^{N}
z_i r_i^{\mu}}{A_1^{\nu_1} \cdots
A_N^{\nu_N}}\right)~,
\label{f16}
\end{equation}
where $z_i~(i=1\cdots N)$ are arbitrary constants and 
$z_0=\sum_{i=1}^{N} z_i$.
After the differentiation and some algebraic manipulations \cite{mi3}
the identity (\ref{f16})
takes the form
\begin{equation}
\sum_{i,j=1}^{N} 
(r_j-r_i)^2 z_i \nu_j I_0^N(D;\{\nu_k+\delta_{k j}\})
=\!\!\!\sum_{i, j=1}^{N}z_i \nu_j
I_0^N(D;\{\nu_k+\delta_{k j}-\delta_{k i}\})
-
(D-\sum_{j=1}^{N}\nu_j) z_0
I_0^N(D;\{\nu_k\}). \label{f20}
\end{equation}
In arriving at (\ref{f20}), it has been understood that
$I_0^N(D;...,
\nu_{l}, 0, \nu_{l+1},...) \equiv$
$I_0^{N-1}(D;..., \nu_{l}, \nu_{l+1},...)$.

The relation (\ref{f20}) represents
the starting point for the derivation 
of the recursion relations for scalar integrals.
We have obtained
the fundamental set of recursion relations 
by choosing the arbitrary constants $z_i~(i=1\cdots N)$
so as to satisfy the following system of linear equations:
\begin{equation}
\sum\nolimits_{i=1}^N 
r_{ij} z_i=C,\quad j=1,\ldots, N,\qquad r_{ij}=(r_i-r_j)^2,\label{f25}
\end{equation} 
where $C$ is an arbitrary constant.
If (\ref{f25}) is taken into account, the relation (\ref{f20}),
after a few manipulations,\cite{mi3} reduces to recursion relation
\begin{equation}
C\,I_0^N(D-2;\{\nu_k\})
=
\sum_{i=1}^{N}z_i I_0^N(D-2;\{\nu_k-\delta_{k i}\})
+(4\pi \mu ^2)
(  D-1-\sum_{j=1}^{N}\nu_j ) z_0
I_0^N(D;\{\nu_k\}), \label{f31} 
\end{equation}
where $z_i$ are given by the solution of the
system (\ref{f25}).
This is a generalized form of the recursion relation
which connects the scalar
integrals in different number of dimensions
\cite{binoth,tarasov,bern}.

By directly choosing the constants $z_i$ in (\ref{f20})
in a such a way that
$z_i=\delta_{i k}$, for $k=1,\cdots, N$, we arrive at
a system of $N$ equations which is always valid:
\begin{equation}
\sum_{j=1}^{N}
(r_k-r_j)^2 \nu_j I_0^N(D;\{\nu_i+\delta_{i j}\})
=\sum_{j=1}^{N} \nu_j
 I_0^N(D;\{\nu_i+\delta_{i j}-\delta_{i k}\})
-(D-\sum_{j=1}^{N}\nu_j)
 I_0^N(D;\{\nu_i\}).\label{f35}
\end{equation}
If the system (\ref{f35}) can be solved with respect to 
$I_0^N(D;\{\nu_i+\delta_{i j}\})$, $j=1,\cdots,N$,
the solutions represent the recursion relations.

The use of the relations (\ref{f31}) and (\ref{f35}) 
in practical calculations
depends on whether the kinematic determinants
\begin{equation}
{\rm det}(R_N)=\left( \begin{array}{ccccc}
 0 & r_{12} & \cdots & r_{1N} \\
 r_{12} & 0 & \cdots & r_{2N} \\
\vdots & \vdots & \ddots & \vdots \\
 r_{1N} & r_{2N} & \cdots & 0
\end{array} \right),
\quad 
{\rm det}(S_N)=\left( \begin{array}{ccccc}
0 & 1 & 1 & \cdots & 1 \\
1 & 0 & r_{12} & \cdots & r_{1N} \\
1 & r_{12} & 0 & \cdots & r_{2N} \\
\vdots & \vdots & \vdots & \ddots & \vdots \\
1 & r_{1N} & r_{2N} & \cdots & 0
\end{array} \right) 
 \label{f32}
\end{equation}  
are equal to zero or not.
We distinguish the following four different types of recursion:
\begin{itemize}
\item Case I: ${\rm det}(S_N)\neq 0$, ${\rm det}(R_N)\neq 0$;
~~($C\neq 0$ and $z_0\neq 0$)
\item Case II: ${\rm det}(S_N)\neq 0$, ${\rm det}(R_N)= 0$;
~~($C=0$ and $z_0\neq 0$)
\item Case III: ${\rm det}(S_N)= 0$, ${\rm det}(R_N)\neq 0$;
~~($C\neq 0$ and $z_0= 0$)
\item Case IV: ${\rm det}(S_N)= 0$, ${\rm det}(R_N)= 0$;
~~($C\neq 0$ and $z_0=0$) or ($C= 0$ and $z_0=0$),
\end{itemize}
where we indicated the necessary choice for the constants $C$ and
$z_0$ in a way that the system (\ref{f25}) has solution and the
most useful recursion relations emerge.

Making use of the relations (\ref{f31}) and (\ref{f35}),
each scalar integral $I_0^N(D;\{\nu_i\})$ can be represented 
as a linear
combination:
\begin{equation}
I_0^N(D;\{\nu_k\})=\sum \nolimits_i c_i(D,r_{ij})\,
 I_0^{N-1}(D^{(i)};\{\nu^{(i)}_k\})+
{\lambda}
\,I_0^N(D';\{1\}),\label{add}
\end{equation}
where for the dimension $D'$ one usually chooses $4+2 \eir$ or
$6+2\eir$. Infinitesimal parameter $\eir$ is regulating the divergences.  
Parameter $\lambda$ equals 0 for Cases II, III and IV \cite{mi3}.
It follows that
in all the above cases, with the exception of the Case I, the integrals with
$N$ external lines can be represented in terms of the 
integrals with smaller number of external lines.
Consequently, then, there exists a fundamental set of
integrals of the form $I_0^N(4+2\eir;\{1\})$ in terms of which all 
integrals can be
represented as a linear combination.

Therefore,\cite{mi3} any 
dimensionally regulated one-loop
N-point Feynman integral can be represented
in terms of six types of box integrals ($N=4$), one type of triangle integral ($N=3$) and
the general (arbitrary $D$, $\nu_1$ and $\nu_2$) two-point integral ($N=2$).
The types of the integrals are determined by number of vanishing kinematic
variables.
Five of the six basic box integrals are IR divergent in 4 dimensions, 
while
the basic triangle integral is finite. However, all the basic box
integrals are finite in 6 dimensions.
Thus, an alternative fundamental set of 
integrals is comprised of five box integrals in 6 dimensions, 
one box and triangle integral in
4 dimensions and the 
general two-point integral. 
This fundamental set of integrals is particularly interesting because
the integral $I_0^2$ is the only divergent
one.
In the final result, 
all the divergences,
IR as well as UV, are contained in the general two-point
integrals and associated coefficients. The expressions for all relevant basic
integrals can be found in the literature \cite{mi3,mi1,mi2}.

\section{Conclusion}

Through the tensor decomposition and scalar reduction
procedure presented, any massless one-loop Feynman integral
with generic 4-dimensional momenta can
be expressed as a linear combination of a fundamental
set of scalar integrals: six box integrals in $D=6$,
a triangle integral in $D=4$, and a general two$-$point
integral. All the divergences present in the original
integral are contained in the general two-point integral
and associated coefficients.

\section*{Acknowledgments}
I would like to thank the organizers of the XXXVIIIth Rencontres de Moriond
conference for the invitation and for an exciting conference.

\end{document}